# Understanding the Interface Dipole of Copper Phthalocyanine (CuPc)/C$_{60}$: Theory and Experiment

Na Sai,[1,*] Raluca Gearba,[1] Andrei Dolocan,[2] John R. Tritsch,[3] Wai-Lun Chan,[1] James R. Chelikowsky,[3] Kevin Leung,[4] Xiaoyang Zhu[1]

[1]Energy Frontier Research Center (EFRC:CST), The University of Texas, Austin, Texas 78712, USA
[2]Texas Materials Institute, The University of Texas, Austin, Texas 78712, USA
[3]Department of Chemistry & Biochemistry, The University of Texas, Austin, Texas 78712, USA
[4]Sandia National Laboratory, MS1415, Albuquerque, New Mexico 87185, USA
* E-mail: nsai@physics.utexas.edu

**Abstract:** Interface dipole determines the electronic energy alignment in donor/acceptor interfaces and plays an important role in organic photovoltaics. Here we present a study combining first principles density functional theory (DFT) with ultraviolet photoemission spectroscopy (UPS) and time-of-flight secondary ion mass spectrometry (TOF-SIMS) to investigate the interface dipole, energy level alignment, and structural properties at the interface between CuPc and C$_{60}$. DFT finds a sizable interface dipole for the face-on orientation, in quantitative agreement with the UPS measurement, and rules out charge transfer as the origin of the interface dipole. Using TOF-SIMS we show that the interfacial morphology for the bilayer CuPc/C$_{60}$ film is characterized by molecular intermixing, containing both the face-on and the edge-on orientation. The complementary experimental and theoretical results provide both insight into the origin of the interface dipole and direct evidence for the effect of interfacial morphology on the interface dipole.

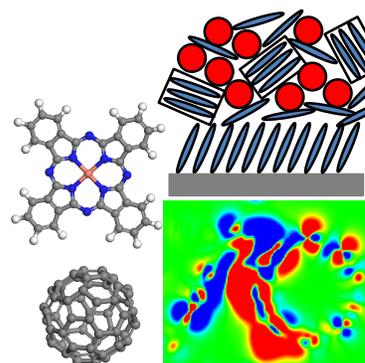

Interfaces between small organic semiconducting molecules and/or polymers play a central role in electronic devices such as organic photovoltaics (OPV) and light emitting diodes.[i] A key issue in the electronic structure of organic interfaces is the donor-acceptor energy level alignment. Despite the weak interaction between organic materials, it has been shown that the simple picture of molecular level alignment with a continuous vacuum level across the interface (Schottky-Mott model) breaks down in a significant number of organic interfaces.[ii] Ultraviolet photoelectron spectroscopic (UPS) studies have reported an interface dipole energy barrier as high as 0.6 eV across many polymer and small molecule interfaces.[2-4] An interface dipole-related electric field identified by vibrational Stark effect spectroscopy has been reported in a polymer blend heterojunction.[5] It has been shown that the interface dipole has important consequences for energy level alignment in organic interfaces[6] and various aspects of OPV performance, including charge separation and recombination rates, and the open circuit voltage.[7]

Despite the importance and the extensive theoretical and experimental studies, the origin of the interface dipole is still subject of debate. Models based on the integer charge transfer[4,8] and the induced density of interface states[9,10] have suggested that the interface dipole is created by charge transfer across the donor/acceptor interface in order to align the polaronic states[4,8] or the charge neutrality levels[9,10] of the two organic materials. On the other hand, recent theoretical studies of the pentacene/$C_{60}$ interface have proposed an explanation based on the polarization effect associated with the anisotropic charge distribution across the interface.[11-12] The presence of metal electrodes and Fermi level pinning between the organic layer and the electrode further complicates the picture for the energy level alignment.[8,13,14] An additional issue surrounding the interface dipole concerns the growth sequence of the donor and acceptor layer. UPS studies of CuPc/$C_{60}$[15] and sexithiophene/$C_{60}$[16] interfaces have reported that the deposition order does not influence the interface dipole. On the other hand, a UPS investigation of pentacene/$C_{60}$[17] has shown a switch in polarity for the interface dipole when the growth sequence is reversed. A theoretical study of the latter system within a self-consistent polarization field theory[11] has attributed the reversing of the interface dipole to a change of molecular orientation. However, a fundamental understanding of the relation between the interfacial morphology and the interface dipole has remained lacking.

UPS spectra in the secondary electron cutoff (SECO) and the valence band regions of a CuPc film, a $C_{60}$ film, and two CuPc/$C_{60}$ interfaces deposited in opposite order on Au (111) are shown in Figure 1a,b. We choose to study this interface because of the prominent value of interface dipole reported in previous UPS studies[15,18-21] and its relevance in photovoltaic applications.[22] The highest occupied molecular orbital (HOMO) offset between CuPc and $C_{60}$ is 1.45 eV for the $C_{60}$ on CuPc film and 1.35 eV for the CuPc on $C_{60}$ film. These values are somewhat higher

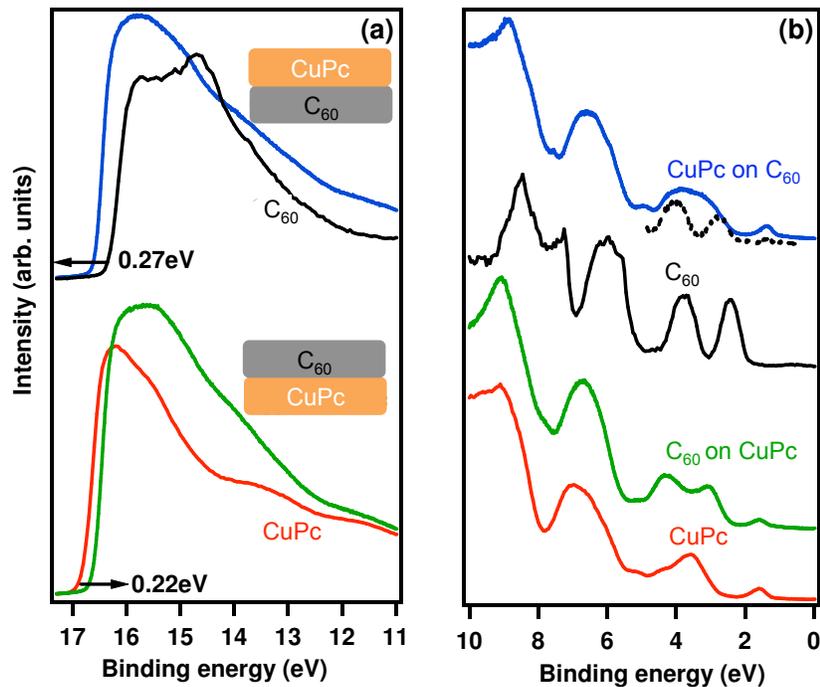

Figure 1. UPS spectra in the (a) SECO and (b) valence band region for the CuPc, $C_{60}$, $C_{60}$ on CuPc, and CuPc on $C_{60}$ films. The dashed line under the spectrum for the CuPc on $C_{60}$ film shows the $C_{60}$ HOMO and HOMO-1 peak resolved by subtracting the CuPc only data. The horizontal axis shows the binding energy relative to the Fermi level

than the 0.95 eV reported by Molodtsova and Knupfer[15] and Zhao and Kahn,[18] but are close to the 1.3 eV and 1.44 eV reported by Brumbach et al.[19] and Akaike et al.[20] Relative to the CuPc film, the SECO of the $C_{60}$ on CuPc film has shifted to the right, giving rise to an upward dipole barrier of +0.22 eV going from CuPc to $C_{60}$ at the interface. Similarly, the SECO of the CuPc on $C_{60}$ film has shifted to the left relative to the $C_{60}$ film, giving rise to a downward dipole barrier of −0.27 eV going from $C_{60}$ to CuPc. Our results for the interface dipole are consistent with those in the literature, which range from 0.28 to 0.5 eV.[15,18-21] The barrier is always positive going from CuPc to $C_{60}$, confirming that the sequence of deposition does not affect the polarity of the interface dipole.

Density functional theory (DFT) calculations based on Hartree−Fock exchange corrected Perdew−Burke−Ernzerhof hybrid functionals were performed on the "edge-on" and "face-on" interfacial systems shown in Figure 2a,b in which the CuPc molecules are standing-up and lying-down with respect to the CuPc (001) and (010) surface, respectively (see details in Computational Methods and Supporting Information). The calculated local densities of states (LDOS) projected onto CuPc and $C_{60}$ together with the calculated band diagram for each interface are shown in Figure 2c,d. The HOMO − HOMO offset between CuPc and $C_{60}$ is calculated to be 1.74 eV for the edge-on interface and 1.14 eV for the face-on interface, while the HOMO to lowest unoccupied molecular orbital (LUMO) band-gaps are 1.72 and 2.3 eV, respectively. The energy offset difference between the lying and standing orientation may be attributed to the potential-energy step intrinsic to the electron distribution of the lying conjugated molecular plane.[23] From the UPS-measured HOMO offset of 1.45 eV and the $C_{60}$ transport gap of 3.5 eV (peak to peak value), we deduce an experimental HOMO − LUMO gap of 2.05 eV, close to the average of the calculated band gaps. Taking into account experimental uncertainties, our calculated band offsets fall in the range of the UPS values and validate the accuracy of the computational approach.

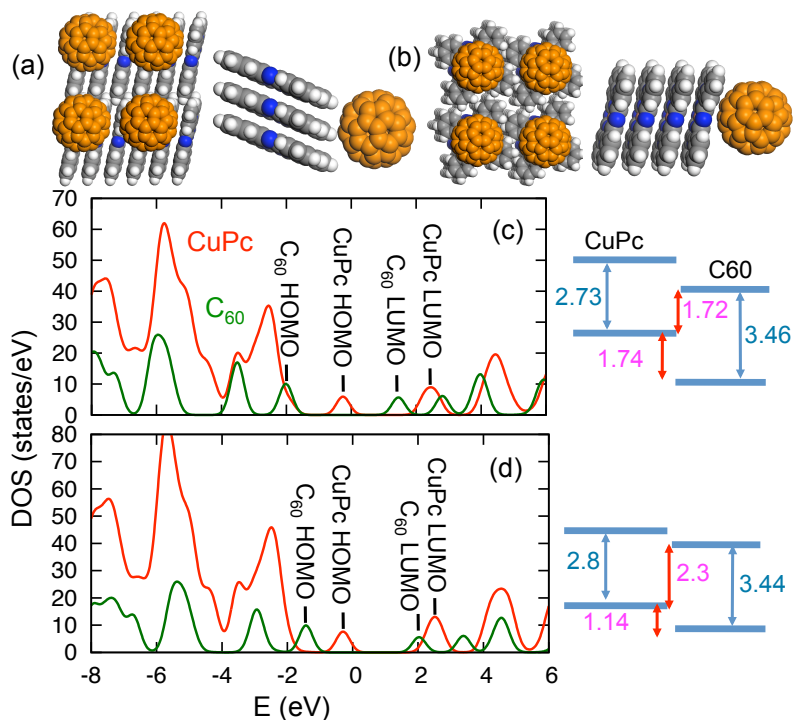

Figure 2. (a), (b) Schematic representation of the edge-on and face-on CuPc/$C_{60}$ interfaces (top and side view). (c,d) Calculated LDOS projected onto CuPc (red) and $C_{60}$ (green) for the edge-on and face-on interface and the corresponding energy level diagrams. The LDOS are convolved with a Gaussian function with 0.2 eV broadening.

To calculate the interface dipole moment, we have integrated the charge density difference $\Delta\rho = \rho_I - (\rho_{CuPc} + \rho_{C60})$, where $\rho_I$, $\rho_{CuPc}$, and $\rho_{C60}$ represent the valence electron densities of the interface, the CuPc slab and the $C_{60}$ layer. Figure 3a,b shows a plane slice of $\Delta\rho$ perpendicular to the interface for the edge-on and face-on orientation and Figure 3c shows the planar average of $\Delta\rho$ along the $z$-direction. The magnitude of the interface dipole moment (μ) for the face-on and edge-on orientation is 0.54 D per $C_{60}$ ($\Delta q = 0.023\ e$) and 0.1 D ($\Delta q = 0.006\ e$), respectively, where $\Delta q$ is the amount of excess (or deficit) of charge obtained by integrating $\Delta\rho$ from the origin to the point where charge depletion switches to charge accumulation. To test the sensitivity of the computed interface dipole to the starting position of $C_{60}$, we have laterally translated the $C_{60}$ molecule along the *b*-axis half way across the supercell and found only a 17% reduction of the interface dipole. We have also increased the thickness of the $C_{60}$ layer to three layers for the face-on interface and found no change in the magnitude of the interface dipole, indicating that the interface dipole is restricted to the interfacial molecules.

The interface dipole induced vacuum level shift ($\Delta\phi$) can be calculated by comparing the electrostatic potential between the interface and the isolated CuPc surface. We have obtained $\Delta\phi$ = 0.13 eV for the face-on interface and 0.04 eV for the edge-on interface. To estimate the interface dipole barrier corresponding to the experimental surface density of $C_{60}$, we have employed the Helmholtz equation $\Delta\phi = \mu\, n\, /\, \varepsilon_0$ and have corrected $\Delta\phi$ by multiplying it with $n_0\, /\, n$, where $n_0 = 0.011$ Å$^{-2}$ is the surface density of $C_{60}$ in the closely packed structure and $n = 1/A$, where $A$ is the unit cell surface area of the model interface. With that, we have obtained $\Delta\phi$ = 0.22 eV for the face-on orientation, which is sizable and in excellent agreement with the UPS data for both the $C_{60}$ on CuPc and CuPc on $C_{60}$ film (0.22 and 0.27 eV respectively). On the other hand, we calculated an interface dipole of $\Delta\phi$ = 0.07 eV for the edge-on orientation that is negligibly small compared to the experimental values, suggesting that the bilayer films in the UPS studies do not exhibit the edge-on configuration at the interface.

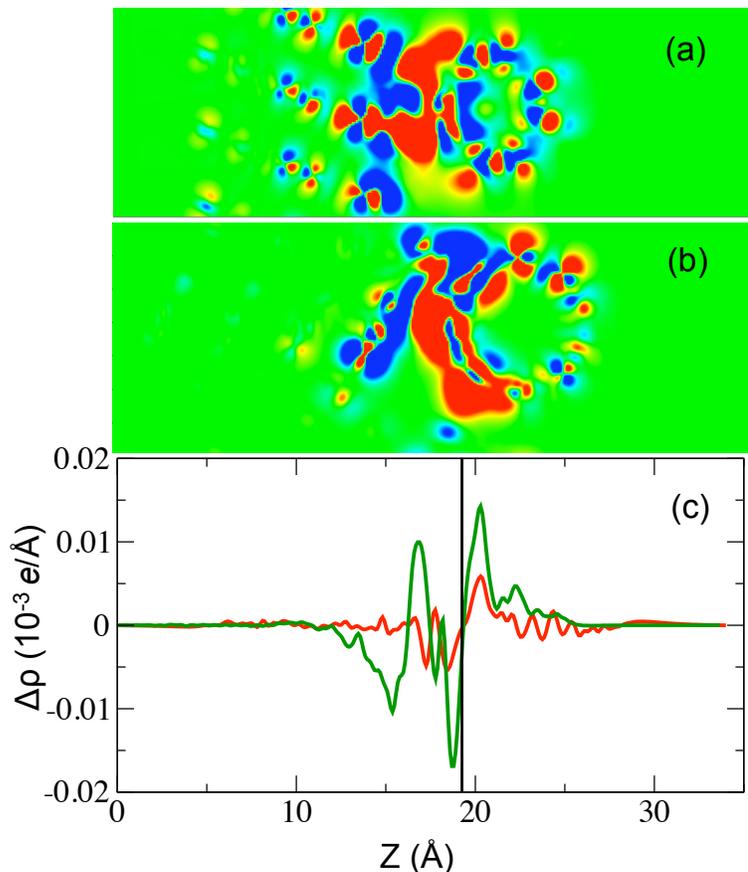

Figure 3. Two-dimensional slice of the charge density difference across the CuPc/$C_{60}$ interface for the (a) edge-on and (b) face-on orientation. (c) Planar averaged charge density difference for the edge-on (red) and face-on (green) orientation as a function of position in the *z*-direction. The vertical line indicates the location where charge depletion becomes charge accumulation.

To understand the origin of the interface dipole, we have applied a Bader charge analysis[24] of the charge density for the face-on interface where the magnitude of the interface dipole is significant. We compare the sum of the Bader charge on the CuPc and $C_{60}$ molecules before and after the formation of the interface. Only a total of ~0.001 $e$ is found to have transferred between CuPc and $C_{60}$,[25] too small to account for the interface dipole moment for the face-on orientation. This result suggests that, despite being widely considered as the mechanism for the observed dipole,[8-10] charge transfer from CuPc to $C_{60}$ is in fact not responsible for the experimentally observed interface dipole barrier. Instead, we believe the interface dipole is a result of charge rearrangement at the interface owing to the polarization effect as suggested by Verlaak et al. and Linares et al.[18,19] The CuPc molecule, similar to benzene or pentacene, has no net dipole but only a quadrupole moment composed of a positive charge on the atoms and a negative charge below and above the ring representing the electron cloud, with an electrostatic potential surface as shown in the Supporting Information. Locally, the negative and positive charges do not cancel and yield an electric field that can repel the electrons away from the CuPc, thus creating an interface dipole. Taking the gradient of the electrostatic potential of a CuPc molecule, we find a stronger electric field perpendicular to the molecular plane than within the plane, consistent with the quantitative difference between the interface dipole for the face-on and the edge-on orientation.

To account for the quantitative agreement between the interface dipoles measured by the UPS and calculated in DFT, we have carried out an interfacial characterization using time of flight-secondary ion mass spectrometry (TOF-SIMS). To our knowledge, this is the first time that the chemical composition of a buried donor/acceptor interface in the direction perpendicular to the interface is reported at a subnanometer resolution. Figure 4 shows the TOF-SIMS depth profiles of $Si/SiO_2/CuPc$ (20 nm), $Si/SiO_2/C_{60}$ (20 nm), $Si/SiO_2/CuPc/C_{60}$(20/20 nm) and $Si/SiO_2/C_{60}/CuPc$ (20/20 nm) films for the $(C_9)^-$ marker. In addition, a wide variety of secondary ion fragments, such as $(C_2N)^-$ that is specific to the CuPc layer only, have been identified and utilized as traces of the interface. The $C_{60}$ on CuPc interface appears as highly intermixed, whereby the $C_{60}$ molecules penetrate into the CuPc layer. The extent of mixing is obtained from the mixing-roughness-information depth (MRI) model.[26] The MRI model is based on the fact that the actual interface is broadened by three phenomenological factors: (1) the roughness of the interface (which can be intrinsic and/or induced by deposition of the second layer and by sputtering), (2) the mixing depth of the sputtering ions, and (3) the secondary ion depth of origin. By using atomic force microscopy, the intrinsic roughness was found to be 1.5 nm in the CuPc films and 0.1 nm in the $C_{60}$ films. The roughness induced by sputtering is neglected (as explained in the Supporting Information), while (2) and (3) give a total length of 1 nm, which was determined from a Au (20 nm)/Cr (10 nm)/Si reference sample.[27] The measured interface length from the depth profile is simply the combination of the above-mentioned parameters (intrinsic corrugation, mixing depth, information depth, and actual interface length) added in quadrature.[28] By fitting a Fermi−Dirac function with a decay time constant $\tau$ to the $C_9^-$ depth profile (at the interface) in Figure 4, we measure an interface length of $(2\ln 3)\tau(R_{C60} + R_{CuPc})$ where $R_{C60}$ and $R_{CuPc}$ are the sputtering rates of the $C_{60}$ and CuPc, respectively, and $\tau$ is the interface sputtering time (see Supporting Information). The sputtering rates for CuPc and $C_{60}$ were determined from the single layer samples to be 0.089 and 0.036 nm/sec, respectively. We obtain a thickness of the mixing layer of 6.5 nm for the $C_{60}$ on CuPc bilayer. The CuPc film grown on a $Si/SiO_2$ substrate by deposition at high temperatures (e.g., 90°C) is polycrystalline[29]. Therefore our films deposited at room temperature are partially amorphous. Similarly to what was reported previously,[30] we assume that the $C_{60}$ diffusion is directed mostly in the amorphous regions of the sample as schematically shown in Figure 4b. Since the lateral resolution is about 1 μm, the TOF-SIMS measures an average of the crystalline and amorphous mixed regions. In these intermixed regions, the CuPc and $C_{60}$ molecules do not form the ideal standing-up configuration as schematically depicted in Figure 4a and observed in thin films far from interfaces,[31] but allow the presence of a face-on in addition to the edge-on orientation. Similarly, in the case of CuPc deposited on top of $C_{60}$, the interface is also characterized by a finite (2.7 nm), albeit smaller amount of mixing (see Figure 4c green curve). The smaller amount of mixing for the case of CuPc deposited on $C_{60}$ can be also inferred from the sputtering rates, which indicate that $C_{60}$ is "harder" than CuPc. The presence of the face-on orientation in the mixed regions in both interfaces explains the quantitative agreement between the DFT prediction and the UPS results. Our results reveal a much more complex interfacial structure than the normally assumed layered models (as shown in Figure 4a). The effect of this complexity must be taken into account to gain a better understanding of the interfacial electronic structure.

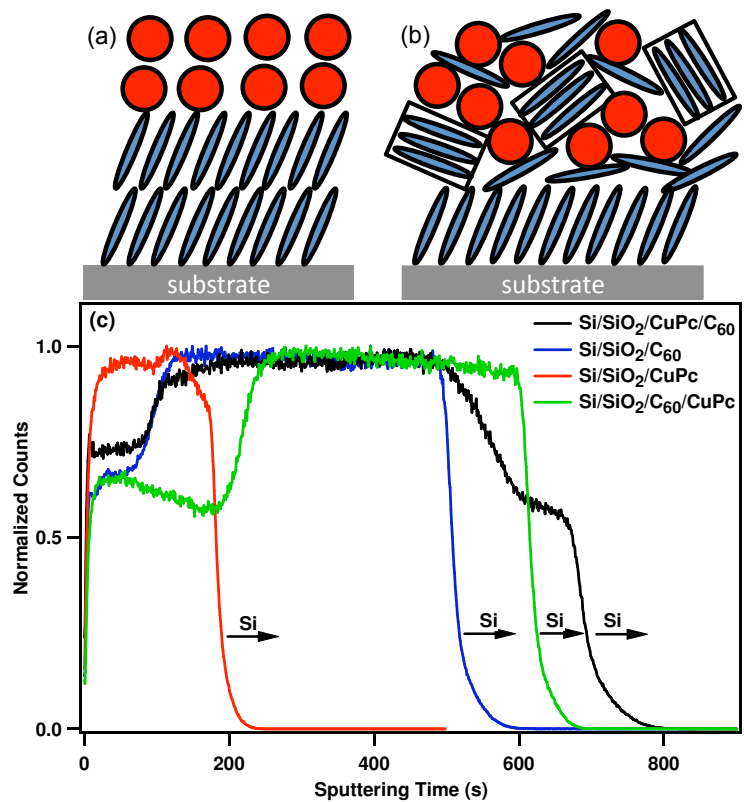

Figure 4. Schematic representation of an ideal standing-up interfacial configuration (a) and molecular intermixed $CuPc/C_{60}$ interface (b). (c) Normalized TOF-SIMS profiles of the $(C9)^-$ marker for the CuPc, $C_{60}$, $C_{60}/CuPc$, and $CuPc/C_{60}$ films on $Si/SiO_2$ substrate.

In summary, combining UPS and *ab-initio* DFT, we have studied the interface dipole barrier for the $CuPc/C_{60}$ interfaces. We find a quantitative agreement between the DFT calculated interface dipole for the face-on orientation and the UPS measured dipole barrier for both the $C_{60}$ on CuPc interface and the CuPc on $C_{60}$ interface. TOF-SIMS shows an intermixed interface, which is consistent with the presence of the face-on orientation and explains the agreement between the dipoles measured in the UPS and the DFT calculations. Our study suggests that a local net charge induced electric field rather than the spontaneous charge transfer across the interface is responsible for the interface dipole. Our study also reveals a complex picture of interfacial morphology. It will be interesting in future work to use tools such as the high-resolution Kelvin probe force microscopy to image the local structure and the spatial profile of the electrostatic potential and thus better understand the effect of the interfacial structures on the interface dipoles.

Eeperimental and Computational Methods

***Density Functional Theory***: The structural relaxation was carried out using PBE exchange-correlation functional with Grimme's van der Waals dispersion correction[32] using the Vienna Ab-initio Simulations Package (VASP).[33] The interface supercell consists of a layer of 3×1 (001) CuPc ($a$ = 12.886 Å, $b$ = 11.307 Å, $c$ = 34.0 Å, $\gamma$ = 90.32°) for the edge-on interface and four layers of CuPc (010) ($a$ =12.06 Å, $b$ = 12.886 Å, $c$ =34.0 Å, $\gamma$ = 90.62°) for the face-on interface, and a vacuum region of about 12 Å. The electronic structure calculations for the interface are carried out using the PBE hybrid functional[34] that mixes 35% Hartree−Fock exchange with the PBE semilocal exchange, a plane-wave energy cutoff of 400 eV, and a Γ-point Brillouin zone sampling.

***Time-of-Flight Secondary Ion Mass Spectrometry:*** The TOF-SIMS data were collected with a TOF-SIMS.5 instrument manufactured by ION-TOF GmbH (Germany, 2010) which provided a mass resolution better than 8000 (m/δm) for all analyzed masses. TOF-SIMS depth profiles were measured using a $Bi_3^+$ primary ion beam (~0.8 pA measured sample current, 30 keV ion energy) for data acquisition and a $Cs^+$ ion beam (~32 nA measured sample current, 500 eV ion energy) for sputtering. The $Bi_3^+$ primary and $Cs^+$ ion beams are raster-scanned over an area of $100 \times 100$ $\mu m^2$ and $250 \times 250$ $\mu m^2$, respectively.

ACKNOWLEDGMENTS
We are grateful to Professors Jean-Luc Brédas, Nobert Koch, and Oliver Monti for helpful discussions and referees for providing comments that helped improve the manuscript. This work was supported as part of the program "Understanding Charge Separation and Transfer at Interfaces in Energy Materials (EFRC:CST)," an Energy Frontier Research Center funded by the U.S. Department of Energy, Office of Science, Office of Basic Energy Sciences under Award Number DE-SC0001091. We acknowledge the use of the TOF-SIMS facility (NSF grant DMR-0923096) of the Texas Materials Institute at The University of Texas at Austin. High performance computing resources used in this work were provided by the National Energy Research Scientific Computing Center and the Texas Advanced Computing Center (TACC). One of us (JRC) would like to acknowledge support from the U.S. Department of Energy grant DE-FG02-06ER46286. Sandia National Laboratories is a multi-program laboratory managed and operated by Sandia Corporation, a wholly owned subsidiary of Lockheed Martin Corporation, for the U.S. Department of Energy's National Nuclear Security Administration under contract DE-AC04-94AL85000.

**Supporting Information**.

Details on the computational and experimental methods and sample preparation. DFT calculated electrostatic potential for the face-on interface and the CuPc molecule. This material is available free of charge via the Internet at http://pubs.acs.org.